\documentclass[a4paper,12pt]{report}
\usepackage{dblfloatfix,fixltx2e,graphicx,array,algorithmic,amssymb,tabularx,cite,url,subfig,algorithm}

\usepackage[cmex10]{amsmath}

\allowdisplaybreaks

\title{Technical report \\   Semisupervised hyperspectral image unmixing using a variational Bayes algorithm\thanks{This research has been co-financed by the European Union (European Social Fund - ESF) and Greek national funds through the Operational Program ”Education and Lifelong Learning” of the National Strategic Reference Framework (NSRF) - Research Funding Program: ARISTEIA- HSI-MARS-1413.}}
\author{K.~Themelis, A.~Rontogiannis,~and~K.~Koutroumbas}
\date{June, 2014}

\begin{document}


\maketitle


\section*{Abstract}
This technical report presents a variational Bayes algorithm for semisupervised hyperspectral image unmixing. The presented Bayesian model employs a heavy tailed, nonnegatively truncated Laplace prior over the abundance coefficients. This prior imposes both the sparsity assumption and the nonnegativity constraint on the abundance coefficients. Experimental results conducted on the Aviris Cuprite data set are presented that demonstrate the effectiveness of the proposed method.

\section*{Introduction }

Hyperspectral remote sensing is a relatively new technology that has gained considerable attention in recent years. It involves the acquisition of image data in many narrow, contiguous spectral bands which provide rich spectral information of the objects imaged. Fig. 1 illustrates the process of generating a pixel's spectral signature out of a hyperspectral image data cube (the cube consists of two spatial and one spectral dimension). The spectral signature of a pixel is simply a vector containing radiance values measured in adjacent spectral bands.  Technological advances in recent years have allowed the implementation of imaging spectrometers which have the ability to collect data in hundreds of adjacent spectral bands. The highly increased volume of data conveys spatial/spectral information that can be properly exploited to accurately determine the type and nature of the objects being imaged. This gives rise to a wide range of applications for hyperspectral image (HSI) processing.

\begin{figure}[h]
\centering
\includegraphics[width=0.9 \linewidth]{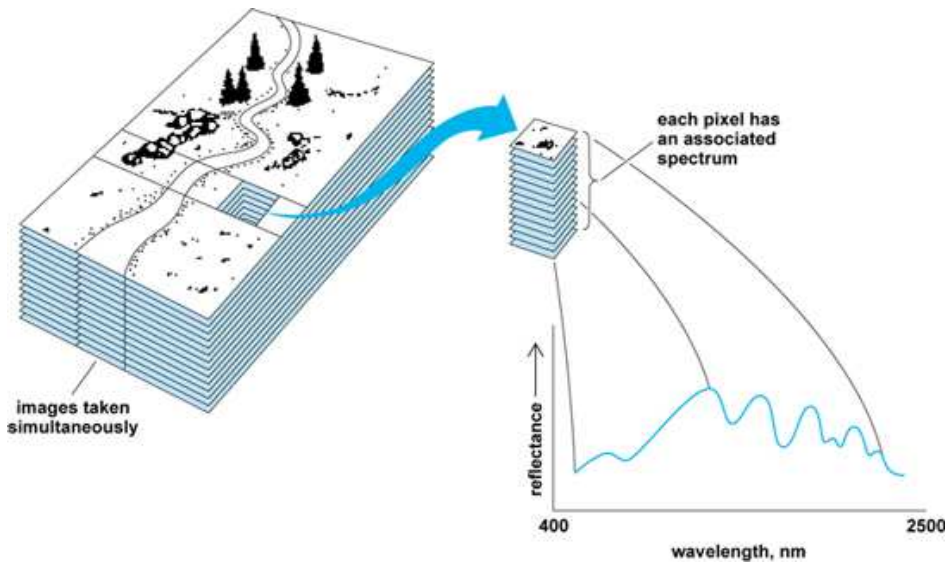}
\caption{Hyperspectral image (modified image taken from \cite{1994_kruse}).}
\label{fig:asbl_mse_vs_iteration_lowpass}
\end{figure}

An intimate limitation of hyperspectral remote sensing is that a single pixel often records a mixed spectral signature of different distinct materials, due to the low spatial resolution of the remote sensor. This raises the need for spectral unmixing (SU), \cite{2003_keshava}, which is a very important step in HSI processing and it has attracted recent interest from the signal and image processing research community. SU is the procedure of decomposing the measured spectrum of an observed pixel into a collection of constituent spectral signatures (or endmembers) and their corresponding proportions (or abundances). A widely used model to perform SU is the linear mixing model. 

Assume a remotely sensed hyperspectral image consisting of $M$ spectral bands, and let ${\mathbf y}$ be a $M\times 1$ vector containing the measured spectral signature (i.e., the radiance values in all spectral bands) of a single pixel. Also let $\boldsymbol \Phi = \left[\boldsymbol\phi_1, \boldsymbol\phi_2, \dots, \boldsymbol\phi_N\right]$ stand for the $M \times N$ endmember signature matrix, where the $M\times 1$ dimensional vector $\boldsymbol\phi_i$ represents the spectral signature of the $i$th endmember, and $N$ is the total number of distinct endmembers present in the scene. Finally, let ${\mathbf w} = \left[w_1, w_2, \dots, w_N \right]^T$ be the $N\times 1$ abundance vector associated with ${\mathbf y}$, where $w_i$ denotes the abundance fraction of $\boldsymbol\phi_i$ in ${\mathbf y}$. The linear mixing model assumes that there is a linear relationship between the spectra of the measured pixel and the 
endmembers, expressed as, 
\begin{align}
{\mathbf y} = \boldsymbol \Phi {\mathbf w} + {\mathbf n}
\label{eq:model_ch2}
\end{align} 
where ${\mathbf n}$ stands for additive noise which is assumed to be a zero-mean Gaussian distributed random vector, with independent and identically distributed (i.i.d.) elements. We write, ${\mathbf n}\sim \mathcal{N}({\mathbf n}|\mathbf{0},\beta^{-1}{I}_{M})$, where $\beta$ denotes the inverse of the noise variance (precision), and ${I}_M$ is the $M \times M$ identity matrix. Two physical constraints are generally imposed into the model described by (\ref{eq:model_ch2}), namely, the abundance non-negativity constraint (ANC), and the abundance sum-to-one (additivity) constraint (ASC), i.e., 
\begin{align}
\mathrm{ANC:}\  w_i \geq 0 ,\ i = 1, 2, \dots, N,\ \mathrm{and }\ \mathrm{ASC:}\ \ \sum_{i=1}^{N} w_i = 1,
\label{eq:constraints}
\end{align}
respectively, although the latter is relaxed in the sequel, e.g., \cite{themelis2012novel}. Utilizing the linear model in (\ref{eq:model_ch2}) and assuming that the endmember matrix $\boldsymbol \Phi$ is known a priori, a constrained linear regression problem is defined, where the parameter of interest is the abundance vector $\mathbf{w}$ for each pixel. 
Additionally, a valid assumption is that only a few of the endmembers present in the image will contribute to the spectrum of a single pixel $\mathbf{y}$. In other words, the abundance vector $\mathbf{w}$ accepts a \emph{sparse} representation in $\boldsymbol \Phi$. Thus, our estimation task consists of estimating $\mathbf{w}$ w.r. to the nonnegativity constraint and the sparsity assumption, given the spectral measurements $\mathbf{y}$ and the endmember matrix $\boldsymbol \Phi$. To this end, we employ the Bayesian framework to define a prior model that expresses our prior knowledge on the parameters of interest, and we then perform Bayesian inference using the variational Bayes algorithm, \cite{jordan1999introduction,attias1999inferring}.

\subsection*{Bayesian modeling}
\label{sec:Bays_model}

The presence of Gaussian noise in (\ref{eq:model_ch2}) dictates that the likelihood function of each pixel's spectral measurement ${\mathbf y}$ is
\begin{align}
p({\mathbf y}|{\mathbf w},\beta) &= \mathcal{N}({\mathbf y}|\boldsymbol \Phi {\mathbf w},\beta^{-1}{\mathbf I}_{M}) \nonumber \\
&= \left( 2\pi\right)^{-\frac{M}{2}} \beta^{\frac{M}{2}}{\rm exp}\left[  -\frac{\beta}{2}{{{\left\| {\mathbf y}  - \boldsymbol \Phi {\mathbf w} \right\|}_2^2}}\right].
\label{eq:likelihood}
\end{align}
The likelihood above is complemented by suitable priors for the model parameters $\{\mathbf{w},\beta\}$. As a prior for the nonnegative noise precision $\beta$ we assume a Gamma distribution, expressed as,
\begin{equation}
p(\beta) = \mathrm{Gamma}(\beta ; \rho,\delta) = \frac{\delta^\rho}{\Gamma(\rho)} \beta^{\rho - 1} \mathrm{exp}\left[ - \delta \beta\right] ,
\label{eq:betaprior}
\end{equation}
where $\rho$ and $\delta$ are its shape and rate parameters, respectively (set to $10^{-6}$ in our experiments). 
For the abundance vector $\mathbf{w}$, we define a two level hierarchical prior that is expressed in a conjugate form and imposes both sparsity and nonnegativity on the abundance coefficients. 
Inspired by \cite{2002_rodriguez}, we select a nonnegatively truncated Gaussian prior for $\mathbf{w}$, i.e.,
\begin{equation}
p(\mathbf{w}| \boldsymbol \alpha,\beta) = \mathcal{N}_{R_{+}^N}\left(\mathbf{w}| \mathbf{0},\beta^{-1} \mathbf{A}^{-1}\right),		
\label{eq:wprior}
\end{equation}
where $\boldsymbol \alpha = [\alpha_1,\alpha_2,\dots,\alpha_N]^\top$ is the precision parameter vector, $\mathbf{A} = \mathrm{diag}(\boldsymbol \alpha)$ is the corresponding diagonal matrix, and $\mathcal{N}_{R_{+}^N}$ signifies the $N$-variate normal distribution truncated at the nonnegative orthant of $R^N$, denoted by $R_{+}^N$, \cite{themelis2012novel}. 
In the second level of hierarchy, the precision parameters $\alpha_i$'s, $i = 1, 2, \dots, N$, are assumed to follow an inverse Gamma distribution, i.e.,
\begin{equation}
p(\alpha_i) = \mathrm{IGamma}(\alpha_i ; 1,\frac{b_i}{2}) = \frac{b_i}{2} \alpha_i^{-2} \mathrm{exp}\left[ - \frac{b_i}{2} \frac{1}{\alpha_i} \right],
\label{eq:alphaprior}
\end{equation}
where $b_i$, $i =1, 2, \dots, N$, is a scale hyperparameter. These two levels of hierarchy form a nonnegatively truncated multivariate Laplace prior over the abundance vector $\mathbf{w}$, which can be established by integrating out the precision $\boldsymbol \alpha$, i.e., 
\begin{align}
& p(\mathbf{w}| \mathbf{b}, \beta) = \int p(\mathbf{w}, \boldsymbol \alpha| \mathbf{b}, \beta) d\boldsymbol \alpha = \int p(\mathbf{w}| \boldsymbol \alpha, \beta) p(\boldsymbol \alpha|\mathbf{b}) d\boldsymbol \alpha  \nonumber \\ 
& = \prod_{i=1}^N \int p(w_i| \alpha_i, \beta) p(\alpha_i|b_i) d\alpha_i \nonumber \\
& = \prod_{i=1}^N \frac{1}{\sqrt{2 \pi}} \sqrt{\beta} \frac{b_i}{2} \int_{0}^{\infty}
 \alpha_i^{-\frac{3}{2}} \mathrm{exp} \left[ - \frac{b_i}{2} \frac{1}{\alpha_i} - \frac{\beta w_i^2 }{2} \alpha_i \right] I_{R_{+}}(w_i) d\alpha_i  \nonumber \\
& = \prod_{i=1}^N \frac{\sqrt{2 \beta} b_i}{\sqrt{\pi}} \left( \frac{\beta w_i^2}{b_i} \right)^{\frac{1}{4}} K_{-\frac{1}{2}}(\sqrt{b_i \beta w_i^2}) I_{R_{+}}(w_i) \nonumber \\
& =  \prod_{i=1}^N \frac{\sqrt{2 \beta} b_i^{\frac{3}{4}}}{\sqrt{\pi}} (\beta w_i^2)^{\frac{1}{4}} \sqrt{\frac{\pi}{2}} \frac{1}{(\beta w_i^2)^{\frac{1}{4}}} \mathrm{exp} \left( - \sqrt{b_i \beta w_i^2}\right)  I_{R_{+}}(w_i)\nonumber \\
& = \prod_{i=1}^N \sqrt{\beta b_i}  \mathrm{exp} \left( - \sqrt{b_i \beta w_i^2}\right) I_{R_{+}}(w_i),
\end{align}
where $I_{R_{+}^N}(\mathbf{w})$ is an indicator function, with $I_{R_{+}^N}(\mathbf{w}) = 1$ (resp. $0$) if $\mathbf{w}  \in R_{+}^N$ (resp. $\mathbf{w} \not \in   R_{+}^N$), and we have used the identities 
\begin{align}
 K_{\frac{1}{2}}(z) = \sqrt{\frac{\pi}{2}} \frac{1}{\sqrt{z}} \mathrm{exp} [-z] ,
\end{align}
and
\begin{align}
 \int_{0}^\infty  x^{\nu-1} \mathrm{exp} \left[ -\frac{\beta}{x} -\gamma x \right] dx = 2 \left( \frac{\beta}{\gamma}\right)^\frac{\nu}{2} K_{\nu}(2 \sqrt{\beta \gamma}). 
\end{align}
%
%
In our formulation, the sparsity-promoting scale hyperparameters $b_i$'s in (\ref{eq:alphaprior}) are also inferred from the data, by assuming the following Gamma prior distribution for each $b_i, i = 1, 2, \dots, N$,   
\begin{equation}
p(b_i) = \mathrm{Gamma}(b_i ; \kappa, \nu) = \frac{\nu^\kappa}{\Gamma(\nu)} b_i^{\kappa - 1} \mathrm{exp}\left[ - \nu b_i \right].
\label{eq:bprior}
\end{equation}
Hyperparameters $\kappa$ and $\nu$ in (\ref{eq:bprior}) are also set to small values ($10^{-6}$ in our experiments). 


\subsection*{Bayesian inference}
\label{sec:var_inf}

The Bayesian approach provides a rigorous way to perform posterior inference for the model parameters through Bayes' rule. The probability density function, $p(\beta, \mathbf{w}, \boldsymbol \alpha, \mathbf{b}|\mathbf{y})$, of our model parameters given the pixel's spectral measurements $\mathbf{y}$ can be expressed using Bayes' rule as
\begin{align}
p(\beta, \mathbf{w}, \boldsymbol \alpha, \mathbf{b}|\mathbf{y}) = \frac{p(\beta, \mathbf{w}, \boldsymbol \alpha, \mathbf{b},\mathbf{y})}{\int p(\beta, \mathbf{w}, \boldsymbol \alpha, \mathbf{b},\mathbf{y}) d\beta d\mathbf{w} d\boldsymbol \alpha d\mathbf{b}}.
 \label{eq:Bayesrule}
\end{align}
However, the integration at the denominator of (\ref{eq:Bayesrule}) is intractable due to the complexity of our model. To circumvent this, we develop a variational Bayes algorithm that approximates the posterior distribution in (\ref{eq:Bayesrule}).

Utilizing the mean field approximation reported in section ???10.9???, we define an approximating distribution, $q(\beta, \mathbf{w}, \boldsymbol \alpha, \mathbf{b})$, which is assumed to factorize as follows,
\begin{align}
q(\beta, \mathbf{w}, \boldsymbol \alpha, \mathbf{b}) = q(\beta) \prod_{i=1}^N q(w_i) \prod_{i=1}^N q(\alpha_i) \prod_{i=1}^N q(b_i).
 \label{eq:meanfield}
\end{align} 
Following the variational Bayes methodology, the individual factors at the right hand side of (\ref{eq:meanfield}) can be computed in closed form.  Note that the conjugacy of our model's prior distributions guarantees that the posterior approximating factors in (\ref{eq:meanfield}) will belong to some known family of probability density functions. Let $\boldsymbol \theta$ be the vector containing all model parameters, i.e., $\boldsymbol \theta = [w_1, \dots, w_N, \beta,$ $ \alpha_1, \dots, \alpha_M, $ $b_1, \dots, b_N]^T$, and $\theta_i$ denote either a $w_j$, or a $\alpha_j$, or a $b_j$, $j=1,\dots, N$, or $\beta$.  
Then, it is known from the variational Bayes theory, \cite{jordan1999introduction}, that 
\begin{align}
q(\theta_i) = \frac{\mathrm{exp} \left( \mathbb{E}_{j \neq i} \left[ \mathrm{log} p(\mathbf{y},\boldsymbol \theta) \right] \right) }{\int \mathrm{exp} \left( \mathbb{E}_{j \neq i} \left[ \mathrm{log} p(\mathbf{y},\boldsymbol \theta) \right] \right) d\theta_i},
\label{eq:gen_qtheta}
\end{align}
where $\mathbb{E}_{j \neq i} \left[ \cdot \right]$ denotes expectation w.r.t. all $q(\theta_j)$'s except for $q(\theta_i)$. Applying (\ref{eq:gen_qtheta}) we compute a nonnegatively truncated Gaussian approximating distribution for each abundance coefficient $w_i$, $i = 1, 2, \dots, N$, i.e., 
\begin{align}
 & q(w_i) = \frac{\mathrm{exp}\left[ \left \langle  \mathrm{log}p(\mathbf{y|\mathbf{w},\beta}) + \mathrm{log}p(\mathbf{w|\boldsymbol \alpha,\beta}) \right \rangle \right]}{\int \mathrm{exp}\left[ \left \langle  \mathrm{log}p(\mathbf{y|\mathbf{w},\beta}) + \mathrm{log}p(\mathbf{w|\boldsymbol \alpha,\beta}) \right \rangle \right]dw_i} \nonumber \\ 
& = \frac{\mathrm{exp} \left[ \left \langle  -\frac{\beta}{2} \|  \mathbf{y} - \boldsymbol \Phi_{\neg i} \mathbf{w}_{\neg i} -  \boldsymbol \phi_i w_i \|^2 - \frac{\beta}{2} \alpha_i w_i^2 + \mathrm{log} I_{R_{+}}(w_i)\right \rangle  \right]}{\int \mathrm{exp} \left[ \left \langle  -\frac{\beta}{2} \|  \mathbf{y} -  \boldsymbol \Phi_{\neg i} \mathbf{w}_{\neg i} -  \boldsymbol \phi_i w_i \|^2 - \frac{\beta}{2} \alpha_i w_i^2 + \mathrm{log} I_{R_{+}}(w_i) \right \rangle \right]dw_i} \nonumber \\
& = \frac{1}{C}  \mathrm{exp} \left[ \left \langle -\frac{\beta}{2} \left( \boldsymbol \phi_i^T  \boldsymbol \phi_i w_i^2 - 2 \boldsymbol \phi_i^T  \left( \mathbf{y} - \boldsymbol \Phi_{\neg i} \mathbf{w}_{\neg i}\right)  w_i + \alpha_i w_i^2 \right) \right \rangle \right] \nonumber \\
& = \frac{1}{C} \mathrm{exp} \left[ \left \langle -\frac{1}{2} \left( \beta \left( \boldsymbol \phi_i^T  \boldsymbol \phi_i + \alpha_i \right) w_i^2  - 2 \beta  \boldsymbol \phi_i^T  \left( \mathbf{y} - \boldsymbol \Phi_{\neg i} \mathbf{w}_{\neg i} \right) w_i \right) \right \rangle \right] \nonumber \\
& = \frac{1}{C} \mathrm{exp} \left[ -\frac{1}{2} \left( \langle \beta \rangle \left( \boldsymbol \phi_i^T  \boldsymbol \phi_i + \langle \alpha_i \rangle \right) w_i^2  - 2 \langle  \beta  \rangle  \boldsymbol \phi_i^T  \left( \mathbf{y} - \boldsymbol \Phi_{\neg i} \langle  \mathbf{w}_{\neg i} \rangle \right) w_i \right) \right]  \nonumber \\
& = \frac{1}{C}(2 \pi)^{-1/2} \sigma_i^{-1} \mathrm{exp} \left[ -\frac{1}{2} \frac{(w_i - \mu_i)^2}{\sigma_i^2}\right] I_{R_{+}}(w_i) \nonumber \\
& = \mathcal{N}_{R_{+}^N}(w_i|\mu_i,\sigma_i^2), 
\label{eq:wpost}
\end{align}
where $\langle \cdot \rangle$ denotes expectation w.r. to the posterior approximating factor $q(\cdot)$, $\boldsymbol \Phi_{\neg i}$ results from $\boldsymbol \Phi$ after removing its $i$-th column, and $\mathbf{w}_{\neg i}$ results from $\mathbf{w}$ after excluding its $i$-th element, $C$ is a normalizing constant, and $\mu_i$ and $\sigma_i^2$ are given by
\begin{align}
& \sigma_i^2 = \langle \beta \rangle^{-1} (\langle \alpha_i \rangle + \boldsymbol \phi_i^T \boldsymbol \phi_i)^{-1} \label{eq:sigmai} \ \mathrm{and}\\ 
& \mu_i = (\langle \alpha_i \rangle + \boldsymbol \phi_i^T \boldsymbol \phi_i)^{-1} \boldsymbol \phi_i^T   (\mathbf{y}- \boldsymbol \Phi_{\neg i} \langle \mathbf{w}_{\neg i} \rangle).
 \label{eq:mui}
\end{align}
The posterior of the precision parameters $\alpha_i$'s is a generalized inverse Gaussian distribution (GIG), computed as 
\begin{align}
 & \mathrm{log} p(\alpha_i) \propto \left \langle \mathrm{log} p(w_i| \alpha_i, \beta)  + \mathrm{log} p(\alpha_i|b_i) \right \rangle = \nonumber \\
 &\left \langle \frac{1}{2} \mathrm{log}\alpha_i - \frac{\beta w_i^2}{2} \alpha_i - 2 \mathrm{log} \alpha_i - \frac{b_i}{2} \frac{1}{\alpha_i} \right \rangle  = \left \langle - \frac{3}{2} \mathrm{log}\alpha_i - \frac{\beta w_i^2}{2} \alpha_i - \frac{b_i}{2} \frac{1}{\alpha_i} \right \rangle  \nonumber \\
 & \Rightarrow  p(\alpha_i) = \frac{ \left( \frac{\langle b_i \rangle}{\langle \beta \rangle \langle w_i^2 \rangle } \right)^{\frac{1}{4}}}{2 K_{1/2}\left(\sqrt{\langle \beta \rangle \langle w_i^2 \rangle \langle b_i \rangle}\right)} \alpha_i^{-\frac{3}{2}} \mathrm{exp}\left[ - \frac{\langle \beta \rangle \langle w_i^2 \rangle}{2} \alpha_i - \frac{\langle b_i \rangle}{2} \frac{1}{\alpha_i}\right] \nonumber \\
 &  = \mathrm{GIG}\left(\alpha_i;\left \langle \beta \right \rangle \left \langle w_i^2 \right \rangle , \left \langle b_i \right \rangle, \frac{1}{2}\right),
 \label{eq:alphaipost}
\end{align}
%
where $K_{\zeta}\left( \cdot \right)$ denotes the modified Bessel function of the second kind with $\zeta$ degrees of freedom. Next, the approximating factors over the hyperparameters $b_i$'s and the noise precision $\beta$ are computed as the following Gamma distributions, i.e.,
\begin{align}
 & \mathrm{log} p(b_i) \propto \left \langle \mathrm{log} p(\alpha_i|b_i) + \mathrm{log} p(b_i)\right \rangle = \nonumber \\
 &\left \langle (\kappa -1 ) \mathrm{log}b_i - \nu b_i + \mathrm{log} b_i - \left \langle \frac{1}{2 \alpha_i}\right \rangle  b_i \right \rangle  = \left \langle \kappa \mathrm{log}b_i  - \left (  \nu + \left \langle \frac{1}{2 \alpha_i}\right \rangle \right) b_i \right \rangle \nonumber \\
 & \Rightarrow p(b_i) = \mathrm{Gamma}\left(b_i;\kappa+1, \nu + \frac{1}{2} \left \langle \frac{1}{ \alpha_i}\right \rangle \right)
 \label{eq:bipost}
\end{align}
%
and 
\begin{align}
 & \mathrm{log} p(\beta) \propto \left \langle \mathrm{log} p(\mathbf{y}| \mathbf{w}, \beta) + \mathrm{log} p(\mathbf{w}| \boldsymbol \alpha, \beta)  + \mathrm{log} p(\beta)\right \rangle = \nonumber \\
 & = \left \langle \frac{M}{2} \mathrm{log} \beta - \frac{\|\mathbf{y} - \boldsymbol \Phi \mathbf{w}\|^2}{2}\beta + \frac{N}{2}\mathrm{log}\beta - \frac{\mathbf{w}^T \mathbf{A} \mathbf{w}}{2}\beta + (\rho -1 )\mathrm{log}\beta - \delta \beta\right \rangle \nonumber \\
 & = \left \langle \left( \frac{M}{2} + \frac{N}{2} + \rho -1 \right) \mathrm{log}\beta  - \left( \frac{\|\mathbf{y} - \boldsymbol \Phi \mathbf{w}\|^2}{2} + \frac{\mathbf{w}^T \mathbf{A} \mathbf{w}}{2} + \delta \right) \beta \right \rangle \nonumber \\
 & \Rightarrow p(\beta) = \mathrm{Gamma}\left(\beta;\frac{M}{2} + \frac{N}{2} + \rho , \frac{ \left \langle \|\mathbf{y} - \boldsymbol \Phi \mathbf{w}\|^2 \right \rangle}{2} + \frac{ \left \langle \mathbf{w}^T \mathbf{A} \mathbf{w} \right \rangle}{2} + \delta  \right)
 \label{eq:betapost}
\end{align}
%
respectively. Notice the interdependency between the parameters of the approximating distributions in (\ref{eq:wpost}), (\ref{eq:alphaipost}), (\ref{eq:bipost}), and (\ref{eq:betapost}). This interdependence gives rise to a cyclic optimization scheme, where, at each step, the expected value of a single parameter (e.g. $\langle w_i \rangle, \langle \beta\rangle$) is updated, while the remaining parameters are kept fixed. In this scheme, the required moments of the model parameters are computed as 
\begin{align}
\langle w_i \rangle  = \mu_i +  \frac{\frac{1}{\sqrt{2\pi}}\mathrm{exp}\left(-\frac{1}{2}\frac{\mu_i^2}{\sigma_i^2}\right)}{1-\frac{1}{2}\mathrm{erfc}\left(\frac{\mu_i}{\sqrt{2}\sigma_{i}}\right)} \sigma_i,
 \label{eq:mui_tr}
\end{align}
\begin{align}
\langle \beta \rangle = \frac{2\rho + M+N}{ 2 \delta + \left \langle \mathbf{w}^T \mathbf{A} \mathbf{w} \right \rangle  + \left \langle \| \mathbf{y} - \boldsymbol \Phi \mathbf{w} \|^2 \right \rangle  } 
\label{eq:betamean}
\end{align}
\begin{align}
\langle \alpha_i \rangle = \sqrt{\frac{\langle b_i \rangle}{\beta \langle w_i^2 \rangle  }}, 
\label{eq:alphaimean}
\end{align}
\begin{align}
\langle b_i \rangle = \frac{\kappa + 1 }{\nu + \frac{1}{2} \left \langle \frac{1}{\alpha_i} \right \rangle}, 
\label{eq:bimean}
\end{align}
where $\mu_i$ and $\sigma_i^2$ are given in (\ref{eq:mui}) and (\ref{eq:sigmai}), respectively, 
\begin{align}
\left \langle \mathbf{w}^T \mathbf{A} \mathbf{w} \right \rangle = \sum_{i=1}^N \langle \alpha_i \rangle \left \langle w_i^2 \right \rangle,
\end{align}
\begin{align}
\left \langle \| \mathbf{y} - \boldsymbol \Phi \mathbf{w} \|^2 \right \rangle = \| \mathbf{y} - \sum_{i=1}^N \boldsymbol \phi_i \langle w_i \rangle  \|^2 +  \sum_{i=1}^N \sigma_{i,tr}^2 \boldsymbol \phi_i^T \boldsymbol \phi_i,  
\end{align}
\begin{align}
\langle w_i^2 \rangle  = \langle w_i \rangle^2 + \sigma_{i,tr}^2, 
\end{align}
\begin{align}
& \sigma_{i,tr}^2 = \sigma_{i}^2\left[1 - \frac{\mu_i}{\sqrt{2\pi} \sigma_i}\frac{\mathrm{exp}\left(-\frac{\mu_i^2}{2\sigma_i^2} \right)}{1 - \frac{1}{2}\mathrm{erfc}\left(\frac{\mu_i}{\sqrt{2}\sigma_i}\right)} - \left(\frac{1}{\sqrt{2\pi}}\frac{\mathrm{exp}\left(-\frac{\mu_i^2}{2\sigma_i^2}\right)}{1 - \frac{1}{2}\mathrm{erfc}\left(\frac{\mu_i}{\sqrt{2}\sigma_i}\right)}\right)^2\right],
\label{eq:sigmai_tr}
\end{align}
and, finally, 
\begin{align}
\left \langle \frac{1}{\alpha_i} \right \rangle = \frac{1}{\langle \alpha_i \rangle} + \frac{1}{\langle b_i \rangle}.
\end{align}
Having expressed all the required moments, our iterative scheme involves updating (\ref{eq:mui_tr}), (\ref{eq:betamean}), (\ref{eq:alphaimean}) and  (\ref{eq:bimean}) in a sequential manner. The resulting variational Bayes algorithm is presented in Algorithm \ref{alg1}. The final estimate on a pixel's abundance coefficient $w_i$ is the mean of the posterior approximating factor $q(w_i), i = 1, 2, \dots, N$.

\begin{algorithm}
\begin{algorithmic}
\STATE Inputs $\mathbf{y}, \boldsymbol \Phi$
\STATE Initialize $\boldsymbol \alpha, \mathbf{w}, \mathbf{b} $
\FOR{$t = 1 , 2, \dots$}
  \STATE $\langle \beta \rangle = (2\rho + M+N)/(2 \delta + \left \langle \mathbf{w}^T \mathbf{A} \mathbf{w} \right \rangle  + \left \langle \| \mathbf{y} - \boldsymbol \Phi \mathbf{w} \|^2 \right \rangle  )$
  \FOR{$i = 1 , 2, \dots, N$}
    \STATE $\langle w_i \rangle  = \mu_i +  \frac{\frac{1}{\sqrt{2\pi}}\mathrm{exp}\left(-\frac{1}{2}\frac{\mu_i^2}{\sigma_i^2}\right)}{1-\frac{1}{2}\mathrm{erfc}\left(\frac{\mu_i}{\sqrt{2}\sigma_{i}}\right)} \sigma_i,$ 
    \STATE $\langle \alpha_i \rangle = \sqrt{\frac{\langle b_i \rangle}{\beta \langle w_i^2 \rangle  }} $
    \STATE $\langle b_i \rangle = \frac{\kappa + 1 }{\nu + \frac{1}{2} \left \langle \frac{1}{\alpha_i} \right \rangle}$
    \ENDFOR
\ENDFOR
\end{algorithmic}
\caption{Proposed variational Bayes scheme}
\label{alg1}
\end{algorithm}

\subsection*{Experimental results }

\begin{figure}[ht]
\centering
\includegraphics[width=0.5 \linewidth]{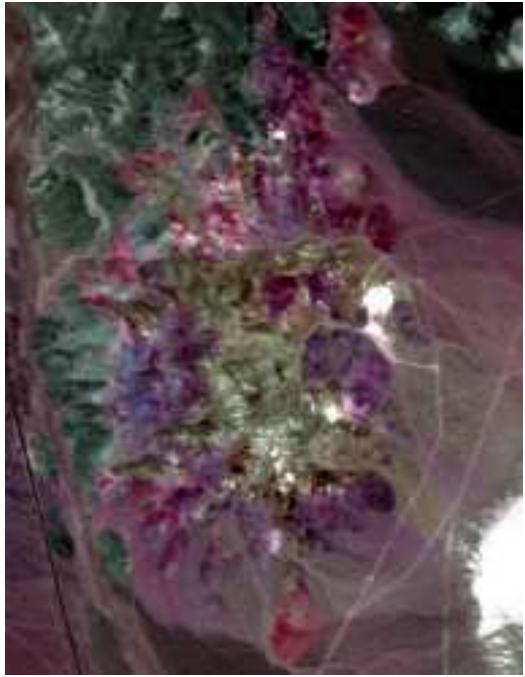}
\caption{RGB composite of the AVIRIS Cuprite subimage using bands $183$, $193$, and $203$. from \cite{1994_kruse}).}
\label{fig:Cuprite_rgb}
\end{figure}

In this section we apply the proposed variational Bayes unmixing algorithm to a real hyperspectral image, collected by the Airborne Visible/Infrared Imaging Spectrometer (AVIRIS) over a Cuprite mining district, in Nevada, in the summer of 1997\footnote{The data are publicly available at \url{http://aviris.jpl.nasa.gov/data/free_data.html}.}. The Cuprite data set has been extensively used to evaluate remote sensing technologies and spectral unmixing algorithms, e.g., \cite{1992_swayze,nascimento2005vertex,miao2007endmember,themelis2012novel,iordache2014collaborative}. It comprises $224$ spectral bands in the range from $400$ to $2500$ nanometers. A subimage of the Cuprite data set with size $250 \times 191$ pixels is used in our experiments. Figure \ref{fig:Cuprite_rgb} displays a pseudocolored composite of our image, where bands $183$, $193$, and $203$ have been used as red, green and blue (RGB) components, respectively. 

\begin{figure}[ht]
\centering
\includegraphics[width=0.75 \linewidth]{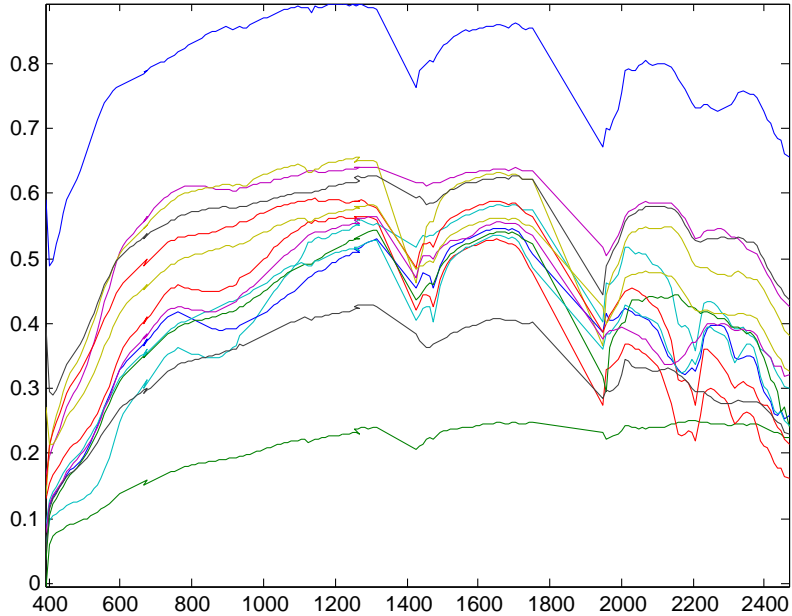}
\caption{Spectral signatures of the $14$ endmembers extracted from the Cuprite image using the VCA algorithm, \cite{nascimento2005vertex}.}
\label{fig:Phi}
\end{figure}

After removing some low SNR bands and water-vapor absorption bands (e.g., bands $1 - 2$, $104 - 113$, $148 - 167$ and $221 - 224$), $188$ spectral bands remain available for processing. As a preprocessing step, we have used the VCA algorithm\footnote{The VCA code is available at \url{http://www.lx.it.pt/~bioucas/code.htm}.}, \cite{nascimento2005vertex}, to extract $14$ endmembers from our hyperspectral image, as in \cite{nascimento2005vertex}. The VCA algorithm identifies the signatures of the ``pure'' pixels in the image and considers them as pure material signatures. A plot of the spectral signatures of the extracted endmembers versus the wavelength is displayed in Figure \ref{fig:Phi}. Notice the high degree of correlation between the spectra of different materials, which is largely responsible for the fact that the obtained endmember matrix $\boldsymbol \Phi$ is ill-conditioned in our inverse problem of abundance estimation. 


\begin{figure}[ht]
\centering
\subfloat[]{\includegraphics[width=0.33\linewidth]{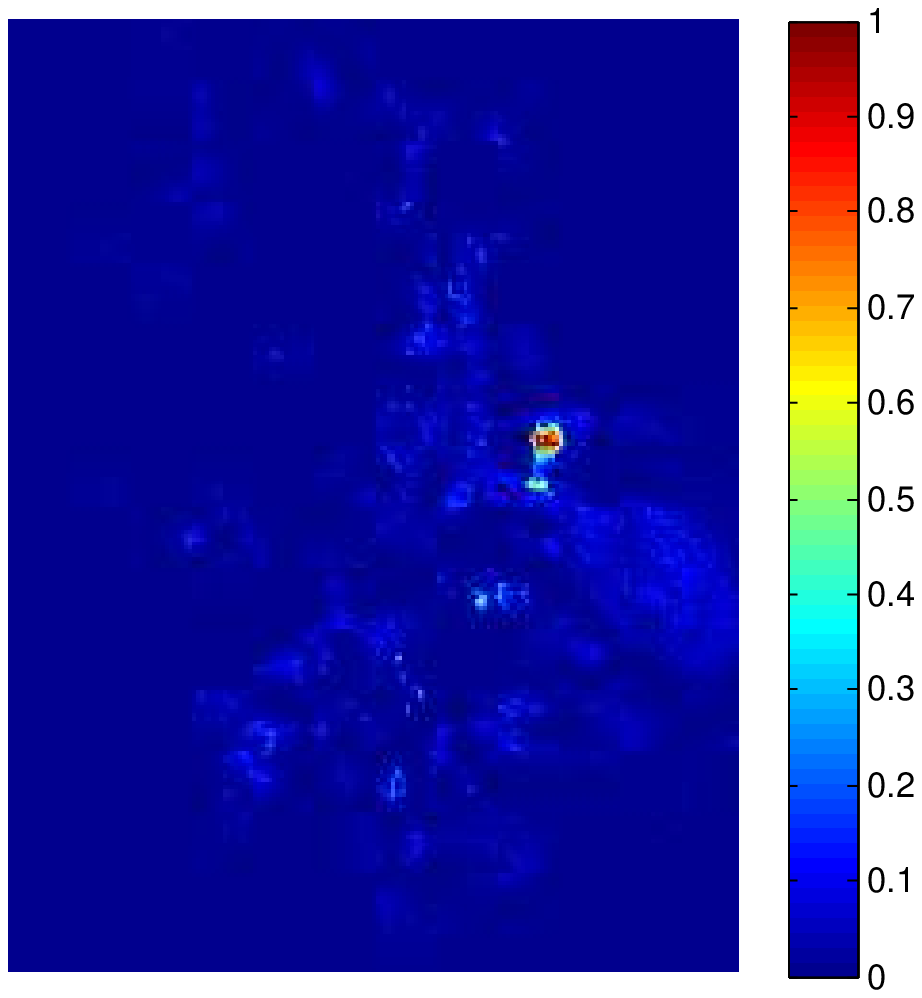}} 
\subfloat[]{\includegraphics[width=0.33\linewidth]{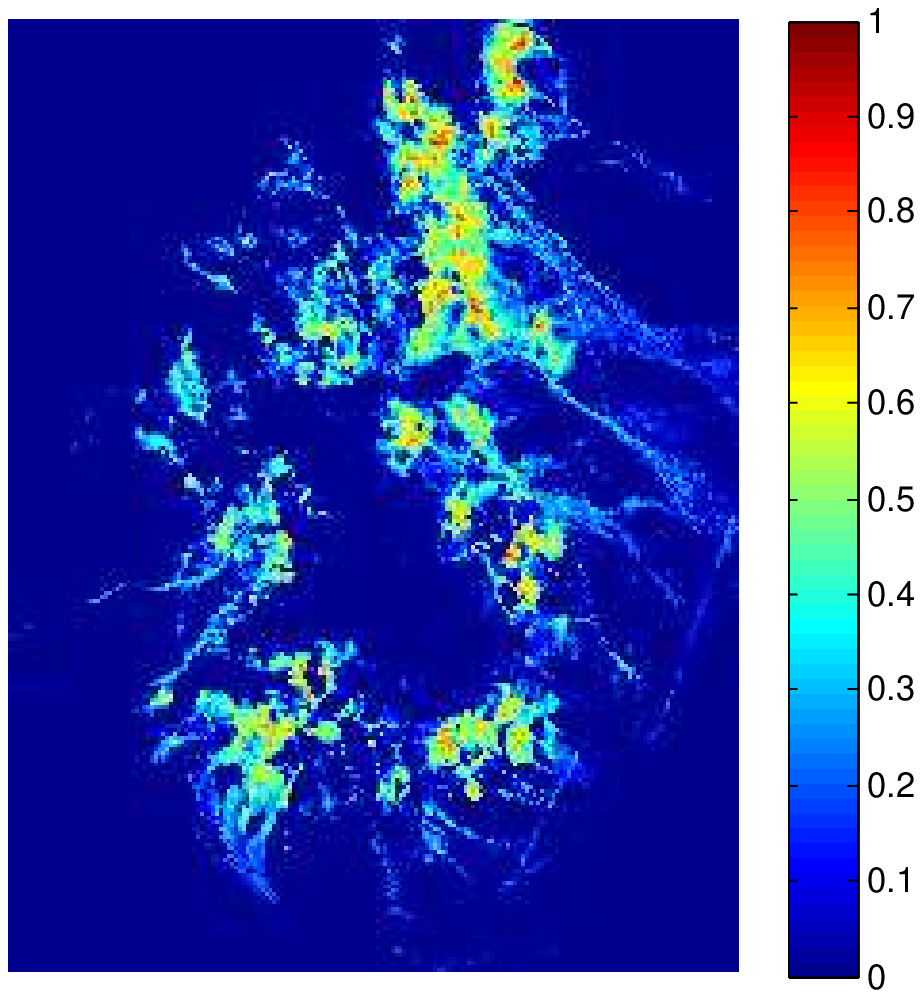}} 
\subfloat[]{\includegraphics[width=0.33\linewidth]{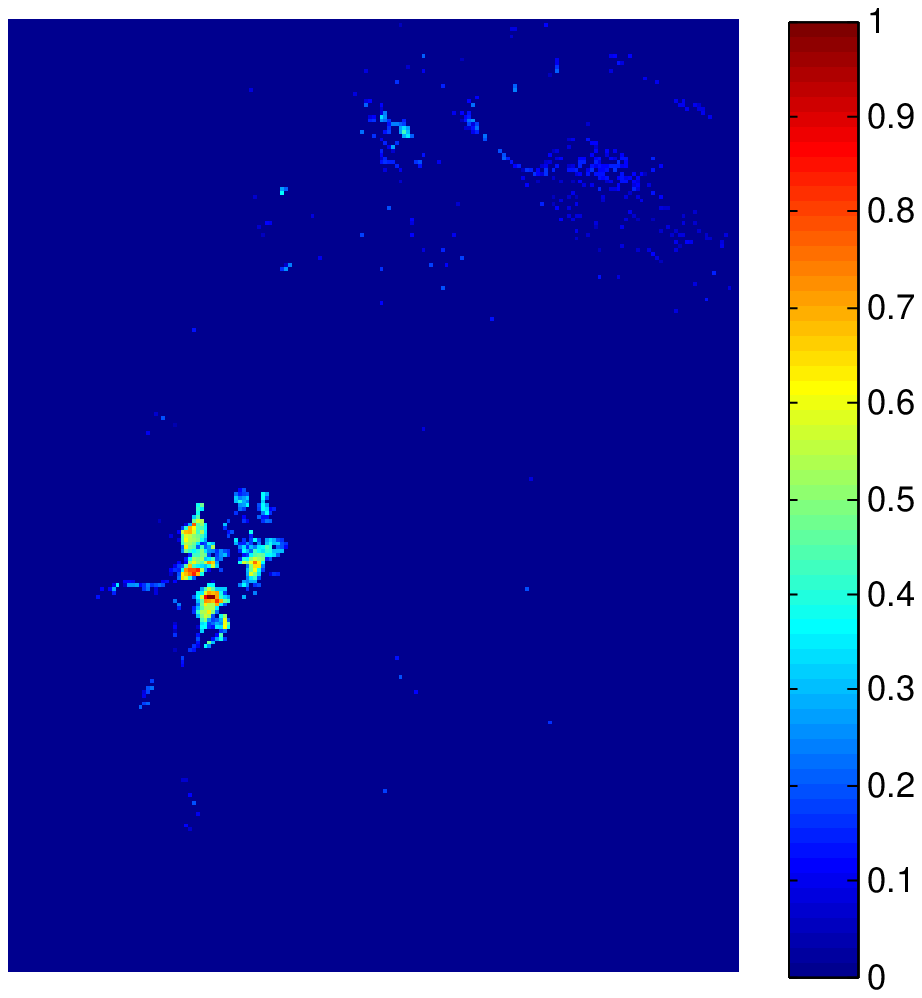}} \hfill
\subfloat[]{\includegraphics[width=0.33\linewidth]{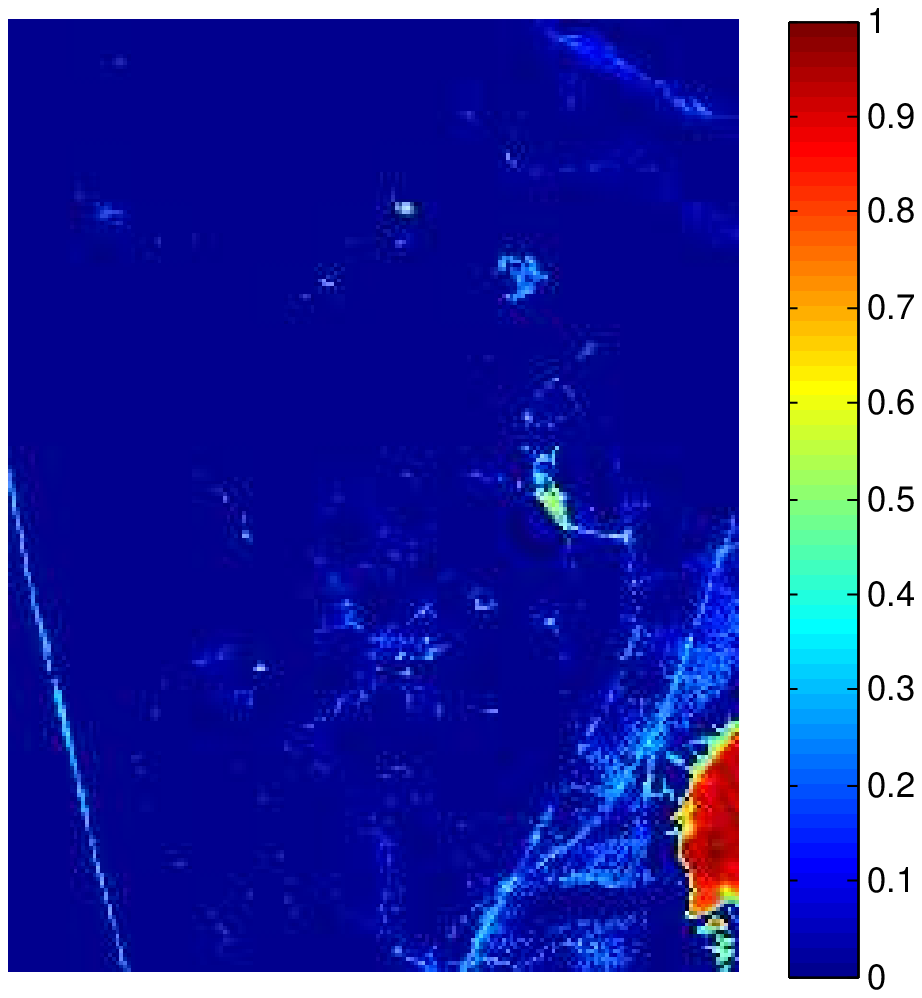}} 
\subfloat[]{\includegraphics[width=0.33\linewidth]{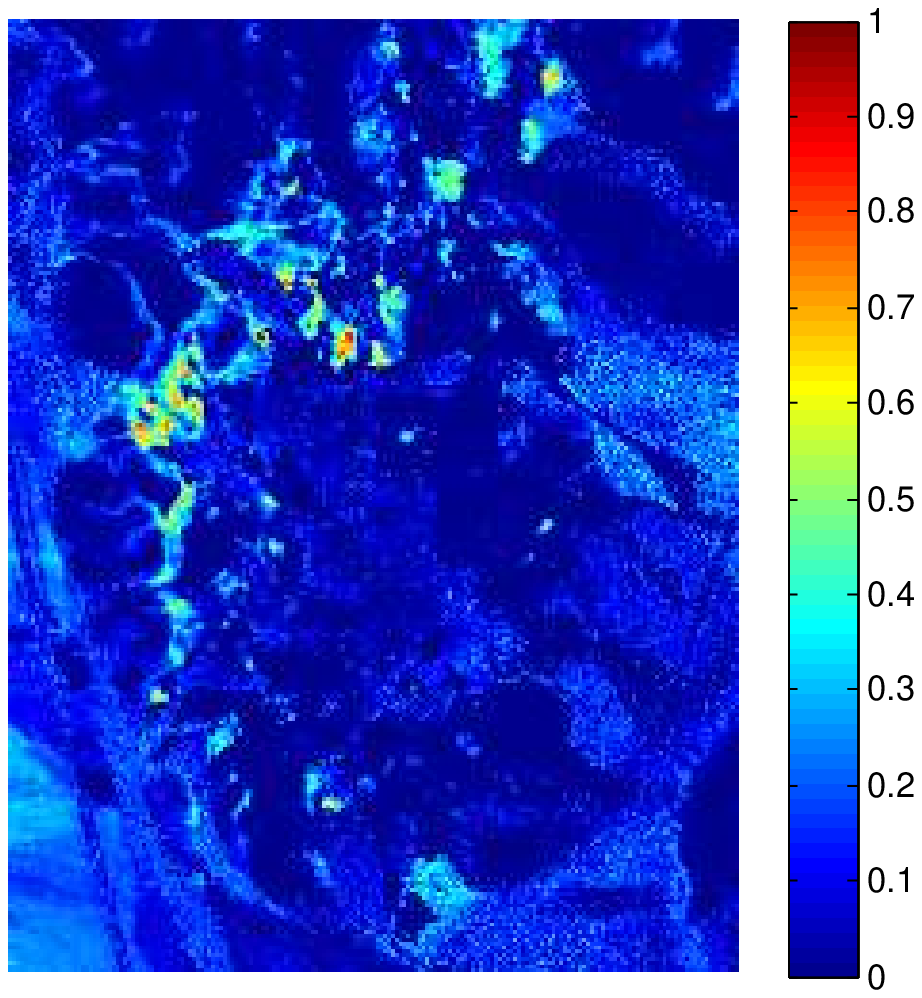}}
\subfloat[]{\includegraphics[width=0.33\linewidth]{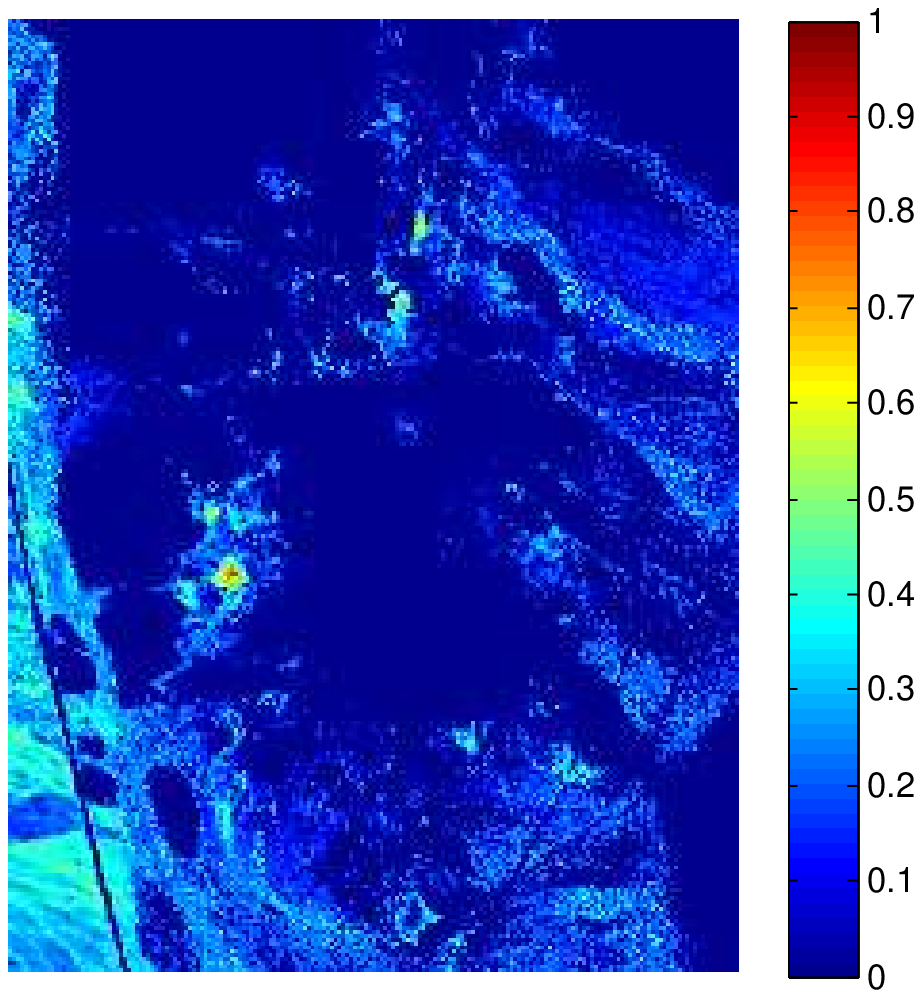}}  
\caption{Estimated abundance maps for the materials (a) Muscovite, (b) Alunite, (c) Buddingtonite, (d) Montmorillonite, (e) Kaolinite 1, and (f) Kaolinite 2, using a variational Bayes algorithm. }
\label{fig:abundance_maps}
\end{figure}

Figure \ref{fig:abundance_maps} shows the resulting abundance maps for six different endmembers using our variational Bayes algorithm. A dark (resp. light) pixel reveals a low (resp. high) proportional percentage for the respective endmember in that pixel. A simple inspection of the abundance maps in Figure \ref{fig:abundance_maps} reveals that they are in accordance with the results in \cite{nascimento2005vertex,miao2007endmember,themelis2012novel,iordache2014collaborative}. More importantly, we are able to identify the presented endmembers in Fig. \ref{fig:abundance_maps} as muscovite, alunite, buddingtonite, montmorillonite, kaolinite 1, kaolinite 2, c.f. \cite{nascimento2005vertex,miao2007endmember,themelis2012novel,iordache2014collaborative}. Note, however, that although our results are quantitatively similar to those presented in the literature, an accurate assessment of our algorithm's estimation performance cannot be established, due to lack of ground truth information. 

 

\bibliographystyle{plain}
\bibliography{refs} 

\end{document}